\documentclass[12pt,twoside]{article}
\usepackage{fleqn,espcrc1}

\usepackage{graphicx}
\usepackage[figuresright]{rotating}

\newcommand{\AmS}{{\protect\the\textfont2
  A\kern-.1667em\lower.5ex\hbox{M}\kern-.125emS}}

\title{Polarized Protons in HERA}

\author{G.~H.~Hoffstaetter\thanks{Invited talk at the NUCLEON99,
                                  INFN, Frascati 1999}
\address{
Deutsches~Elektronen-Synchrotron DESY,
        Notkestrasse~85, D-22603~Hamburg, FRG}}
\begin{document}

\maketitle


\begin{abstract}
  Polarized proton beams at HERA can currently only be produced by
  extracting a beam from a polarized source and then accelerating it
  in the three synchrotrons at DESY.  In this paper, the processes
  which can depolarize a proton beam in circular accelerators are
  explained, devices which could avoid this depolarization in the DESY
  accelerator chain are described, and specific problems which become
  important at the high energies of HERA are mentioned.  At HERA's
  high energies, spin motion cannot be accurately described with the
  isolated resonance model which has been successfully used for lower
  energy rings.  To illustrate the principles of more accurate
  simulations, the invariant spin field is introduced to describe the
  equilibrium polarization state of a beam and the changes during
  acceleration. It will be shown how linearized spin motion leads to a
  computationally quick approximation for the invariant spin field and
  how to amend this with more time consuming but accurate
  non-perturbative computations.  Analysis with these techniques has
  allowed us to establish optimal Siberian Snake schemes for HERA.
\end{abstract}

\section{INTRODUCTION}

In contrast to polarized high energy electron beams which can become
polarized by the emission of spin flip synchrotron radiation, proton
beams do not become polarized after acceleration.  High energy
polarized protons can only be produced by accelerating a beam from a
polarized ion source.  The highest energy reached so far was 25GeV in
the AGS and a lot of care had to be taken during acceleration to
preserve this injected polarization \cite{roser98c,roser99a}.
To explain depolarization in circular accelerators, some concepts from
spin dynamics have to be introduced.
When a particle with charge $q$ moves through a magnetic field, the
motion of the classical spin vector in the instantaneous rest frame is
described by the Lorentz force equation and the Thomas-BMT equation
\cite{thomas27,bargmann59},
\begin{equation}
\frac{d\vec p}{dt}=-\frac{q}{m\gamma}\{\vec B_\perp\}\times\vec p\ ,\
\ \ \ \ \frac{d\vec s}{dt}=-\frac{q}{m\gamma}\{(G\gamma+1)\vec B_\perp
+(G+1)\vec B_\parallel\}\times\vec s\ ,
\label{eq:bmt}
\end{equation}
where $\vec B_\perp$ and $\vec B_\perp$ are the magnetic field
components perpendicular and parallel to the particle's momentum $\vec
p\;$.  At high energy where $(G\gamma+1)/\gamma\approx G$, the spin
motion is independent of energy and a fixed field integral of $5.48$Tm
leads to a spin rotation of $180^\circ\;$, in contrast to the orbit
deflection which varies with $1/\gamma$.  For fixed orbit deflections
and thus fixed ratio of $\vec B_\perp/\gamma$, the spin precession
rate increases with energy as we now describe.

In purely transverse magnetic fields, the Thomas-BMT equation has the
same structure as the Lorentz force equation up to a factor
$G\gamma+1\;$.  The spin therefore rotates like the momentum but with
a magnified rate.  At the HERA energy of 920GeV, the magnification
factor is $G\gamma=1756\;$ so that the spin is rotated by
$100^\circ\;$ when a proton's direction is altered by 1mrad in a
transverse magnetic field and the spin rotates 1756 times around the
vertical while a particle makes one turn on the design orbit of a flat
circular accelerator, where the fields are vertical. The number of
spin rotations which a particle performs while it travels along the
closed orbit once is called the spin tune $\nu_0\;$.

When the spin tune is integer, the spin comes back to a field
imperfection with the same direction after one turn and the effect of
the field error can add up coherently from turn to turn.  This
resonant depolarization at integer spin tunes $\nu_0$ is called an
{\em imperfection resonance} \cite{courant80}.
  
When viewed at a fixed azimuth $\theta$ of the accelerator, the
particles appear to perform harmonic oscillations around the closed
orbit with the frequencies $\nu_x$, $\nu_y$, and $\nu_\tau$ for
horizontal, vertical, and longitudinal motion. These are called the
orbital tunes.  Some of the fields through which a particle propagates
will therefore oscillate with the orbital tunes.  Whenever the
non-integer part of the spin tune is equal to plus or minus one of
these frequencies, the resulting coherent perturbation can lead to
depolarization.  The coherent depolarization at the first order
resonance condition $\nu_0=m+\nu_k$ is called an {\em intrinsic
  resonance} \cite{courant80}.  Here the notation $\nu_1$=$\nu_x$,
$\nu_2$=-$\nu_x$, $\nu_3$=$\nu_y$, $\nu_4$=-$\nu_y$,
$\nu_5$=$\nu_\tau$, $\nu_6$=-$\nu_\tau$ is used.  Since the spin tune
changes with energy (in a flat ring $\nu_0=G\gamma$) resonances
will have to be crossed at some energies during acceleration.

After one turn around the accelerator, all spins of particles on the
closed orbit have been rotated by $2\pi\nu_0$ around a unit rotation
vector $\vec n_0\;$.  This vector is determined by the accelerator's
main guide fields and small field imperfections only perturb $\vec
n_0$ weakly except at energies where the guide fields would produce an
integer spin tune.  Then, when viewed from a fixed azimuth, spins
would come back after one turn apparently without a rotation.  Close
to imperfection resonances the remaining rotation and therefore the
direction of $\vec n_0$ will be dominated by the influence of field
errors.  While the energy changes during acceleration, $\vec n_0$
changes its direction strongly at these resonances.  Whenever this
change is sufficiently slow, spins which are initially parallel to
$\vec n_0$ will follow the change of $\vec n_0$ adiabatically.
Imperfection resonances can therefore be crossed either by making the
field imperfections small enough or by making them so strong that
$\vec n_0$ already starts to get influenced by the field errors
sufficiently long before the resonance and then changes slowly enough
to let all spins follow adiabatically while the resonance is crossed.
Special magnets for enhancing this effect without disturbing the orbit
are referred to as partial snakes.  So far solenoid magnets have been
used \cite{krisch94} but for the AGS a helical dipole partial snake is
under construction \cite{roser99a}.

The motion of spins along phase space trajectories is dominated by the
main guide fields on the closed orbit except close to an intrinsic
resonance, where the coherent perturbations described above can
dominate over the main guide fields.  When the emittance of the beam
and therefore the amplitude of the perturbations is sufficiently
small, intrinsic resonances can be crossed without loss of
polarization.  Polarization in the core of the beam will therefore be
only weakly influenced when crossing intrinsic resonances.  If a
strong coherent perturbation is slowly switched on and off, an effect
similar to adiabatically following $\vec n_0$ occurs and polarization
is conserved.  Therefore, while an intrinsic resonance is crossed,
perturbations influencing particles in the tails of a beam will slowly
increase already before the resonance and this adiabatic conservation
of polarization can occur.  In intermediate parts of the beam,
however, the polarization is lost.  This type of depolarization can be
overcome by slowly exciting the whole beam coherently at a frequency
close to the orbital tune which causes the perturbation.  All spins
then follow the adiabatic change of the polarization direction and the
resonance can be crossed with little loss of polarization.  The
excitation amplitude is then reduced slowly so that the beam emittance
does not change noticeably during the whole process.  This mechanism
has recently been tested successfully at the AGS \cite{roser98c}.  An
older technique of avoiding depolarization at strong intrinsic
resonances utilizes pulsed quadrupoles to move the orbital tune within
a few microseconds just before a resonance so that the resonance is
crossed so quickly that the spin motion is hardly disturbed
\cite{krisch89}.

So far no polarized beam has been accelerated to more than 25GeV
\cite{roser99a}.  But the possibility of polarized proton acceleration
has been analyzed for RHIC (250GeV), for the TEVATRON (900GeV), and
for HERA (920GeV).  When accelerating through approximately 5000
resonances in the case of HERA, even very small depolarization in
every resonance crossing would add up to a detrimental effect.

It was mentioned below equation (\ref{eq:bmt}) that in a fixed
transverse magnetic field the deflection angle of high energy
particles depends on energy, whereas the spin rotation does not depend
on energy.  It is therefore possible to devise a fixed field magnetic
device which rotates spins by $\pi$ whenever a high energy particle
travels through it at the different energies of an acceleration cycle.
Such field arrangements which rotate spins by $\pi$ while perturbing
the orbit only moderately are called Siberian Snakes
\cite{derbenev73}.  The rotation axis is called the snake axis and the
angle of this axis to the beam direction is referred to as the snake
angle $\psi\;$.  Let us consider a Siberian Snake with snake angle
$\psi_1$ at one point in a flat ring and a second Siberian Snake with
snake angle $\psi_2$ at the opposite side of the ring where the spin
has rotated by $G\gamma/2$.  The spin rotation around the vertical
between the Siberian Snakes is described with Pauli matrices by the
quaternion $\cos(\pi G\gamma/2)+i\sin(\pi G\gamma/2)\sigma_2$.  The
rotation by the first Siberian Snake is described by
$i[\sin(\psi_1)\sigma_1+\cos(\psi_1)\sigma_3]\;$.  The total rotation
for one turn around the ring is then described by
\begin{eqnarray}
&&\phantom{\cdot}\;
i[\sin(\psi_1)\sigma_1+\cos(\psi_1)\sigma_3]\cdot
 [\cos(\pi G\gamma/2)+i\sin(\pi G\gamma/2)\sigma_2]
\nonumber\\
&&\cdot\;
i[\sin(\psi_2)\sigma_1+\cos(\psi_2)\sigma_3]\cdot
 [\cos(\pi G\gamma/2)+i\sin(\pi G\gamma/2)\sigma_2]
\nonumber\\
&=&\phantom{\cdot}\;
i[\sin(\psi_1+\pi G\gamma/2)\sigma_1+\cos(\psi_1+\pi G\gamma/2)\sigma_3]
\nonumber\\
&&\cdot\;
i[\sin(\psi_2+\pi G\gamma/2)\sigma_1+\cos(\psi_2+\pi G\gamma/2)\sigma_3]
\nonumber\\
&=&
-\cos(\psi_1-\psi_2)+i\sin(\psi_1-\psi_2)\sigma_2\ .
\end{eqnarray}
For $\psi_1-\psi_2=\pi/2$ the spins rotate in total $1/2$ times around
the vertical $\vec n_0$ during a complete turn around the ring, giving
$\nu_0=1/2$.  All imperfection resonances and, since the orbital tunes
cannot be $1/2$, also all first order intrinsic resonances are avoided
by the insertion of these two Siberian Snakes, and polarized beam
acceleration to very high energy could become possible.
Siberian Snakes can only be used at sufficiently high energies since
their fields are not changed during acceleration of the beam and they
produce orbit distortions which are too big for energies below
approximately 8GeV$\;$ \cite{anferov99a}.

\section{THE DESY ACCELERATOR CHAIN FOR POLARIZED PROTONS}

For HERA a polarized proton beam would be produced by a polarized
H${}^-$ source.  Then it would be accelerated to 750keV in an RFQ and
then to 50MeV in the LINAC~III from where it would be accelerated in
the synchrotron DESY~III to 7.5GeV/c.  In the next ring, PETRA,
40GeV/c are reached, and HERA finally accelerates to 920GeV/c.
The
four main challenges for obtaining highly polarized beams in HERA are:
(1) {\em Production} of a 20mA pulsed H${}^-$ beam.  (2) {\em
  Polarimetry} at various stages in the acceleration chain.  (3) {\em
  Acceleration} through the complete accelerator chain with little
depolarization.  (4) {\em Storage} of a polarized beam at the top
energy over many hours with little depolarization.

Polarized protons are produced either by a polarized atomic beam
source (ABS), where a pulsed beam with 87\% polarization for 1mA beam
current has been achieved, or by an optically pumped polarized ion
source (OPPIS), where pulsed beams with 60\% for 5mA have been
achieved.  Experts claim that 80\% polarization and 20mA could be
achievable with the second type of source.  The current source at DESY
produces 60mA but the maximal current of 205mA in DESY~III can
already be achieved with a 20mA source.

Polarimeters will have to be installed at several crucial places in
the accelerator chain.  The source would contain a Lyman-$\alpha$
polarimeter \cite{zelenski86}.  Another polarimeter could be installed after
the RFQ \cite{buchmann91}.  This could not be operated continuously since it
disturbs the beam.  The transfer of polarized particles through the
LINAC~III could be optimized with a polarimeter similar to that in the
AGS LINAC; and like the AGS, DESY~III could contain an
internal polarimeter \cite{krisch89}.  Polarization at DESY~III energies
has been achieved and measured at several labs already.  It is
different with PETRA and HERA energies; for these high energies there
is no established polarimeter.  Here one has to wait and see how the
novel techniques envisaged and developed for RHIC will work
\cite{bunce99krisch99}.

Since DESY~III has a super period of eight, only 4 strong intrinsic
first order resonances have to be crossed.  They are at values for the
spin tune $G\gamma$ of $8-\nu_y$, $0+\nu_y$, $16-\nu_y$, and
$8+\nu_y$.  Depolarization can be avoided by jumping the tune with
pulsed quadrupoles in a few microseconds or by excitation of a
resonance with an RF dipole.  A solenoid partial snake would be used
to cross the one strong imperfection resonance at $G\gamma=8$.  All
these methods have been tested successfully at the AGS and it is
likely that a highly polarized proton beam could be extracted from the
DESY~III synchrotron at 7.5GeV/c$\;$.

In PETRA it would be very cumbersome to cross all resonances which can
be seen in figure \ref{fig:res}~(middle).  Since Siberian Snakes can
be constructed for the injection energy of PETRA
\cite{hoff96g} it will be best to avoid all first order
resonances with two such devices.  There is space for Siberian Snakes
in the east and the west section of PETRA.

\section{SPECIFIC PROBLEMS FOR THE HERA RING}

HERA is a very complex accelerator and a brief look already indicates
four reasons why producing a polarized beam in HERA is more difficult
than in an ideal ring.  (1) HERA has a super periodicity of one and
only an approximate mirror symmetry between the North and South halves
of the ring.  Therefore more resonances appear than in a ring with
some higher super periodicity and special schemes for canceling
resonances in symmetric lattices \cite{hoff99b2} cannot be
used in such a ring.  (2) The proton ring of HERA is on top of the
electron ring in the arcs, and the proton beam is bent down to the
level of the electron ring on both sides of the three experiments H1,
HERMES, and ZEUS in the North, East, and South straight sections.  The
HERA proton accelerator is therefore not a flat ring.  The destructive
effect of the vertical bends can, however, be eliminated by so called
flattening snakes \cite{steffen88,anferov97} which let the spin motion
in pairs of vertical bends cancel and makes $\vec n_0$ vertical
outside the non-flat sections of HERA.  (3) There is space for spin
rotators which make the polarization parallel to the beam direction
inside the collider experiments while keeping it vertical in the arcs,
and there is also space for four Siberian Snakes.  But installing more
than four Siberian Snakes would involve a lot of costly construction
work.  Simulations have shown that 8 snakes with properly chosen snake
angles would be desirable.  However, if one does not choose optimal
snake angles, then four-snake-schemes can be better than eight snake
schemes \cite{hoff99b1}.  (4) The energy is very high and
therefore the spin rotates rapidly.  If HERA had been designed for
polarized proton acceleration, several parts of the ring would
probably have been constructed differently.

\section{APPLICABLE THEORY AND SIMULATION TOOLS}

\subsection{The isolated resonance model}

In the isolated resonance model, the field components which perturb
the spin of a particle that oscillates around the closed orbit are
Fourier expanded.  The perturbation of spin motion is then
approximated by dropping all except one of the Fourier components.
When $\vec z$ describes the phase space coordinates relative to the
closed orbit and $\theta$ describes the accelerator's azimuth, the
Thomas-BMT equation (\ref{eq:bmt}) has the form $d\vec
s/d\theta=\vec\Omega(\vec z,\theta)\times\vec s$.  The precession
vector $\vec\Omega$ can be written as
$\vec\Omega_0(\theta)+\vec\omega(\vec z,\theta)$ with a part on the
closed orbit and a part which is linear in the phase space coordinates
$\vec z$.  For spins parallel to the rotation vector on the closed
orbit $\vec n_0(\theta)$ only the components of $\vec\omega(\vec
z,\theta)$ which are perpendicular to $\vec n_0$ perturb the
polarization.  We now choose two mutually orthogonal unit vectors
$\vec m_0$ and $\vec l_0$ which are perpendicular to $\vec n_0$ and
precess around $\vec\Omega_0$ according to the Thomas-BMT equation on
the closed orbit.  The frequency of their rotation is given by the
spin tune $\nu_0$.

In this model a depolarizing resonance occures when a Fourier
component of $\vec\omega(\vec z(\theta),\theta)$ rotates with the same
frequency as $\vec m_0$ and $\vec l_0$ so that there is a coherent
perturbation of the spins away from $\vec n_0$.  The Fourier component
$\epsilon_{\nu_0}$ for this frequency is obtained from the Fourier
integral along a trajectory $\vec z(\theta)$,
\begin{equation}
\epsilon_{\nu_0}=\lim_{N\to\infty}\frac{1}{2\pi N}\int_0^{2\pi N}
\vec\omega(\vec z(\theta),\theta)\cdot(\vec m_0+i\vec l_0) d\theta\ .
\label{eq:res}
\end{equation}
These resonance strengths are shown in the figure \ref{fig:res}~(top),
(middle), and (bottom) for the three proton synchrotrons at DESY.
They were all computed for an oscillation amplitude of $\vec
z(\theta)$ corresponding to the one sigma vertical emittance of
$4\pi$mm$\;$mrad.

\subsection{The invariant spin field}

Already at extraction from PETRA the polarized beam would have
somewhat more energy than any other polarized proton beam so far
obtained and one has to ask whether the isolated resonance model
successfully used so far for describing depolarization is still
applicable.  To understand whether the isolated resonance model
describes spin motion at HERA accurately, we introduce the invariant
spin field of a circular accelerator.  It has been mentioned that a
particle on the closed orbit has to be polarized parallel to $\vec
n_0$ in order to have the same polarization after every turn.
Similarly, one can ask if the whole field of spin directions for
particles at different phase space points can be invariant from turn
to turn.

Each particle can have a different spin direction at its phase space
point $\vec z\;$ and each of these spins propagates with a different
precession vector $\vec\Omega(\vec z(\theta),\theta)$ in the
Thomas-BMT equation.  A spin field $\vec n(\vec z)$ which is invariant
after one turn around the ring is called an invariant spin field or a
Derbenev--Kontratenko $\vec n$-axis \cite{derbenev73}.  A beam which
is polarized parallel to this invariant spin field at every phase
space point does not change its polarization state from turn to turn.
Particles change their location in phase space from some initial phase
space coordinate $\vec z_i$ in the Poincar\'e section at azimuth
$\theta$ to some final coordinate after one turn $\vec z_f=\vec M(\vec
z_i)$ according to the one turn map.  And spins change their
directions according to the one turn spin transport matrix $\underline
R(\vec z_i)$, but the invariant field of spin directions $\vec n(\vec
z_i)$ does not change after one turn.  This requirement is encompassed
by the periodicity condition
\begin{equation}
\vec n(\vec M(\vec z_i))=\underline R(\vec z_i)\vec n(\vec z_i)\ .
\end{equation}
Note that the polarization state of a particle beam is in general not
invariant from turn to turn when all particles are initially
completely polarized parallel to each other, but rather when each
particle is polarized parallel to $\vec n(\vec z)$ at its phase space
point $\vec z$.  In this case the polarization of a particle will be
parallel to $\vec n(\vec z_i)$ whenever it comes close to its initial
phase space point $\vec z_i$ during later turns around the ring, as
long as $\vec n(\vec z)$ is sufficiently continuous.  When two
particles travel along the same trajectory, the angle between their
two spins does not change.  When a particle is initially polarized
with an angle $\phi$ to $\vec n(\vec z)$, it will therefore be rotated
around $\vec n(\vec z)$ every time it comes close to $\vec z_i$, but
it will still have the angle $\phi$ to the invariant spin field.  The
time averaged polarization at $\vec z_i$ will therefore be parallel to
$\vec n(\vec z_i)$, but it can only have the magnitude 1 if the spin
was initially parallel to the invariant spin field.  However, even if
all particles are initially polarized parallel to $\vec n(\vec z)$,
the beam polarization is not 1 but $<\vec n>$ where $<\ldots>$ denotes
an average over the beam.  The maximum average polarization that can
be stored in an accelerator at a given fixed energy is therefore
$|<\vec n>|$.  It was first pointed out in \cite{barber98c} that this
maximum polarization can be small in HERA.

Since the spin dynamics depends on energy, the invariant spin field
$\vec n(\vec z)$ will change during the acceleration process. If this
change is slow enough, spins which are parallel to $\vec n(\vec z)$
will follow adiabatically.  However, if the change is too rapid,
polarization will be lost.  It is therefore good to have $<\vec n>$
close to 1 not only at the collider energy but during the complete
acceleration cycle.  Four problems occur when the different directions
of $\vec n(\vec z)$ are not close to parallel for all particles in the
beam.  (1) Sudden changes of $\vec n(\vec z)$ reduces the
polarization.  (2) The average polarization available to the collider
experiment is reduced.  (3) The polarization involved in each
collision process depends on the phase space position of the
interacting particles. (4) Measuring the polarization in the tail of
the beam will not give accurate information on the average
polarization of the beam.

\subsubsection{Linearized spin orbit motion}

At azimuth $\theta$, a spin can be described by a usually small
complex coordinate $\alpha$ with $\vec s=\Re\{\alpha\}\vec
m_0(\theta)+\Im\{\alpha\}\vec l_0(\theta)+\sqrt{1-|\alpha|^2}\vec
n_0(\theta)$.  When the spin coordinates $\alpha$ and the phase space
coordinates are linearized, one approximates an
initial spin by $\vec s_i\approx\Re\{\alpha_i\}\vec
m_0(0)+\Im\{\alpha_i\}\vec l_0(0)+\vec n_0(0)$ at azimuth $0$ and the
final spin after one turn around the accelerator by $\vec
s_f=\Re\{\alpha_f\}\vec m_0(0)+\Im\{\alpha_f\}\vec l_0(0)+\vec
n_0(0)\;$ where $\alpha_f$ is determined by the $7\times 7$ one turn
transport matrix $\underline M_{77}$,
\begin{equation}
{\vec z_f\choose\alpha_f} =
\underline M_{77}{\vec z_i\choose\alpha_i}=
{\underline M\ \ \ \ \vec 0\choose{\vec G}^T\ e^{i2\pi\nu_0}}
{\vec z_i\choose\alpha_i}\ ,
\end{equation}
whereby $\underline M$ is the $6\times 6$ dimensional one turn transport
matrix for the phase space variables, the exponential describes the
rotation of the spin components $\alpha$ by the spin tune $\nu_0$
around $\vec n_0$, the row vector ${\vec G}^T$ describes the
dependence of spin motion on phase space motion to first order, and
the 6 dimensional zero vector $\vec 0$ shows that the effect of Stern
Gerlach forces on the orbit motion is not considered.

We now write the components perpendicular to $\vec n_0$ of the
invariant spin field as a complex function $n_{\alpha}(\vec z)$ and
use a 7 dimensional vector $\vec n_1$ to obtain the first order
expansion of $\vec n(\vec z)$.  The linearized periodicity condition
for the invariant spin field is
\begin{equation}
\vec n_1(\vec z)={\vec z\choose n_\alpha(\vec z)}\ ,\ \
\vec n_1(\underline M\vec z)=\underline M_{77}\vec n_1(\vec z)\ .
\label{eq:period1}
\end{equation}
This equation can be solved for $\vec n_1$ after the matrices are
diagonalized.  Let $\underline A^{-1}$ be the column matrix of
eigenvectors of the one turn matrix $\underline M$.  The diagonalized
matrix of orbit motion $\underline\Lambda=\underline A\;\underline
M\;\underline A^{-1}$ has the diagonal elements $\exp(i2\pi\nu_k)$
given by the orbital tunes $\nu_1=\nu_x$, $\nu_2=-\nu_x$, etc.  We now
need the $7\times 6$ dimensional matrix $\underline T$ which is the
column matrix of the first 6 eigenvectors of $\underline M_{77}$ and
has the form
\begin{equation}
\underline T={\underline A^{-1}\choose{\vec B}^T}\ , \ \
\underline T\;\underline\Lambda=\underline M_{77}\underline T\ ,
\label{eq:eigen}
\end{equation}
where the 7th components of the eigenvectors form a vector $\vec B$.
If a linear function $\vec n_1(\vec z)=\underline K\vec z$ of the
phase space coordinates can be found, which satisfies the periodicity
condition (\ref{eq:period1}), then an invariant spin field has been
determined.  Inserting the form $\vec z_1=\underline K\vec z$ into
equation (\ref{eq:period1}) and multiplying the resulting condition
$\underline K\;\underline M=\underline M_{77}\underline K$ by
$\underline A^{-1}$ from the right leads to $\underline K\;\underline
A^{-1}\underline\Lambda = \underline M_{77}\underline K\underline
A^{-1}\;$. Therefore $\underline K\;\underline A^{-1}$ is the $7\times
6$ dimensional matrix of eigenvectors $\underline T$ satisfying
equation (\ref{eq:eigen}) and we conclude that there exists a unique
linear invariant spin field given by
\begin{equation}
\vec n_1(\vec z)=\underline T\;\underline A\vec z\ .
\end{equation}
In the linear approximation of spin motion, the
invariant spin field is simply computed via the eigenvectors of the
$7\times 7$ spin orbit transport matrix.  This matrix $\underline
M_{77}$ can be computed in various ways, for example by multiplying
the individual spin transport matrixes of all elements \cite{chao81b}
or by concatenating spin transport quaternions of individual elements
as done in the program SPRINT \cite{hoff96d}.  In the normal
form space belonging to the diagonal matrix $\underline\Lambda$ the
coordinates are given by the actions $J_j$ and the angle variables
$\Phi_j$ with
\begin{equation}
\underline A\vec z =
{{\sqrt{J_1}e^{i\Phi_1}\atop\sqrt{J_1}e^{-i\Phi_1}}\choose\ldots}\ .
\end{equation}
The average over all angle variables of a phase space torus then leads
to the average opening angle of
\begin{equation}
<\phi(\vec n,\vec n_0)>
\approx
{\rm atan}(\sqrt{<|n_\alpha|^2>})
=
{\rm atan}(\sqrt{\sum_{k=1}^3(|B_{2k-1}|^2+|B_{2k}|^2)J_k})\ ,
\end{equation}
where the $B_{k}$ are the 7th components of the eigenvectors
in equation (\ref{eq:eigen}).

\begin{sidewaysfigure}[htb]
\begin{minipage}[t]{110mm}
\includegraphics[width=110mm,bb=129 590 460 717]{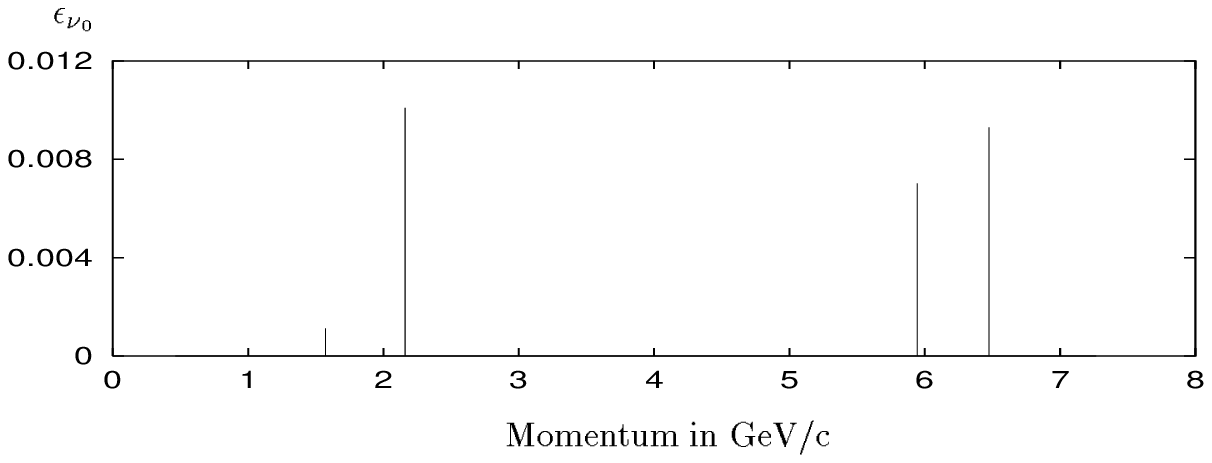}
\includegraphics[width=110mm,bb=129 590 460 717]{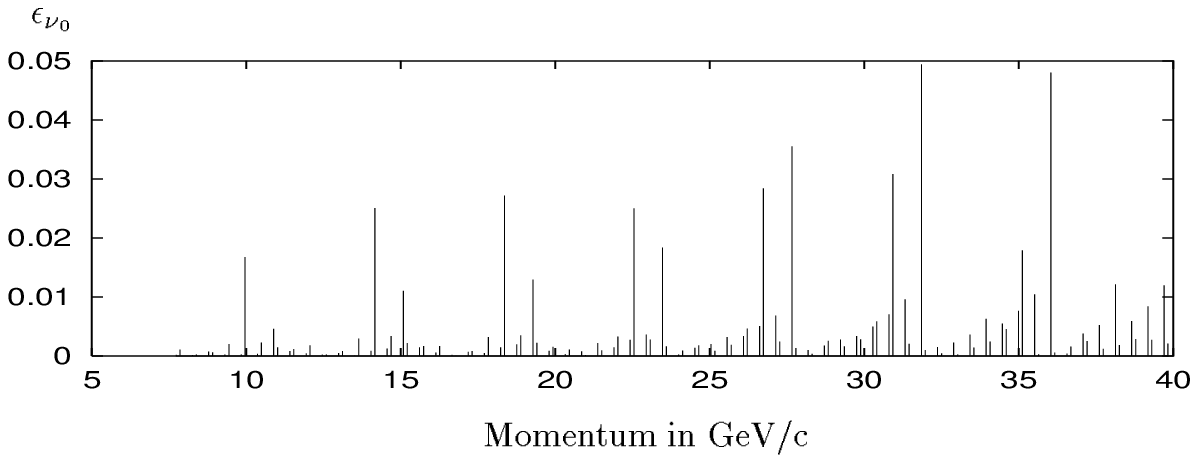}
\includegraphics[width=110mm,bb=129 590 460 717]{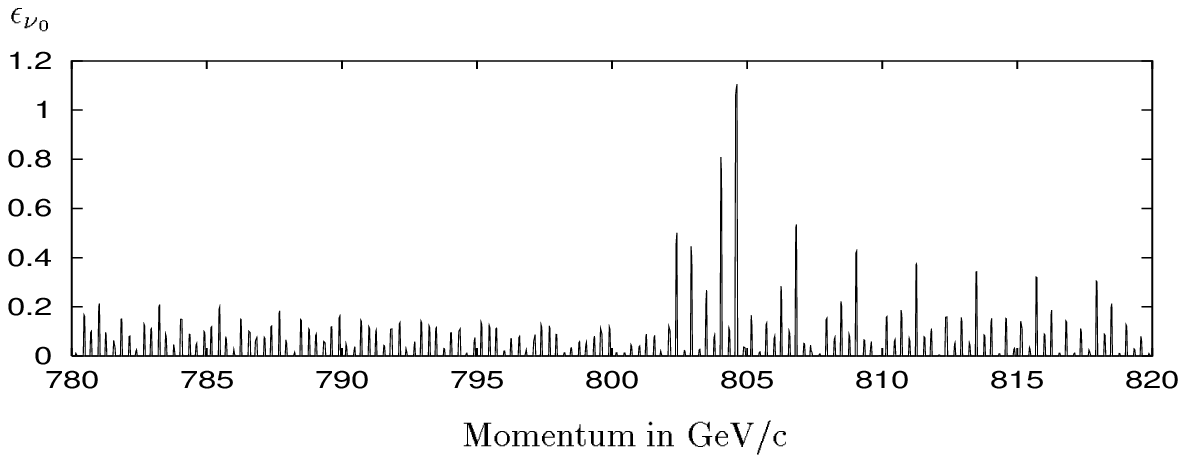}
\caption{Resonance strength for DESY~III, PETRA, and HERA.
\label{fig:res}}
\end{minipage}
\hfill
\begin{minipage}[t]{110mm}
\includegraphics[width=110mm,bb=129 590 460 717]{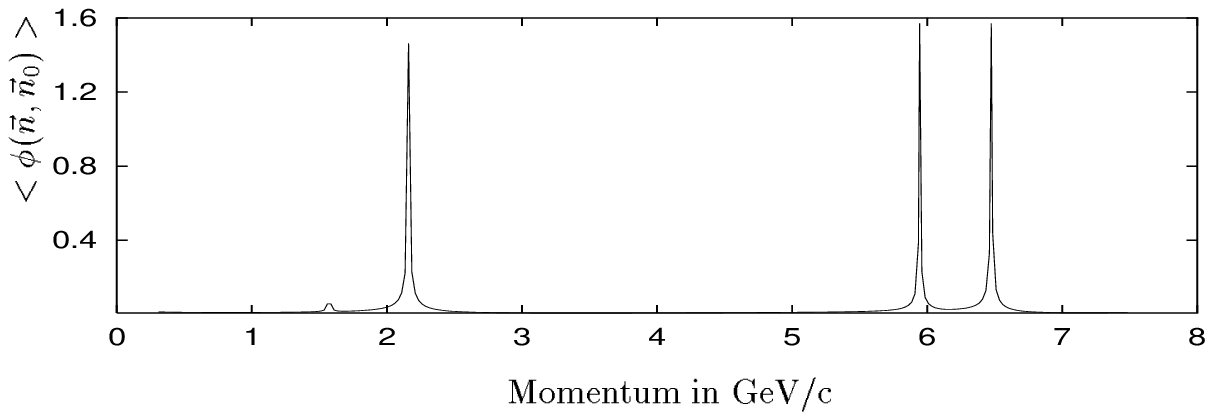}
\includegraphics[width=110mm,bb=129 590 460 717]{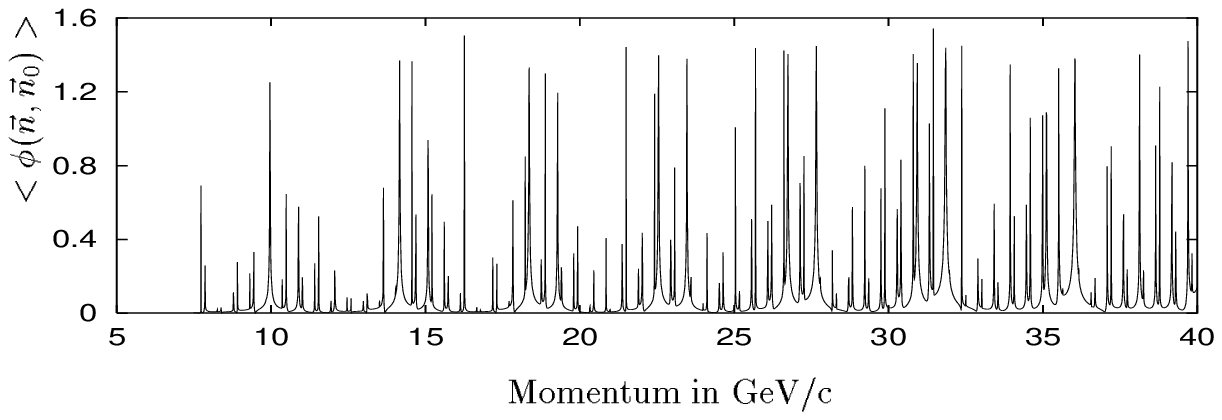}
\includegraphics[width=110mm,bb=129 590 460 717]{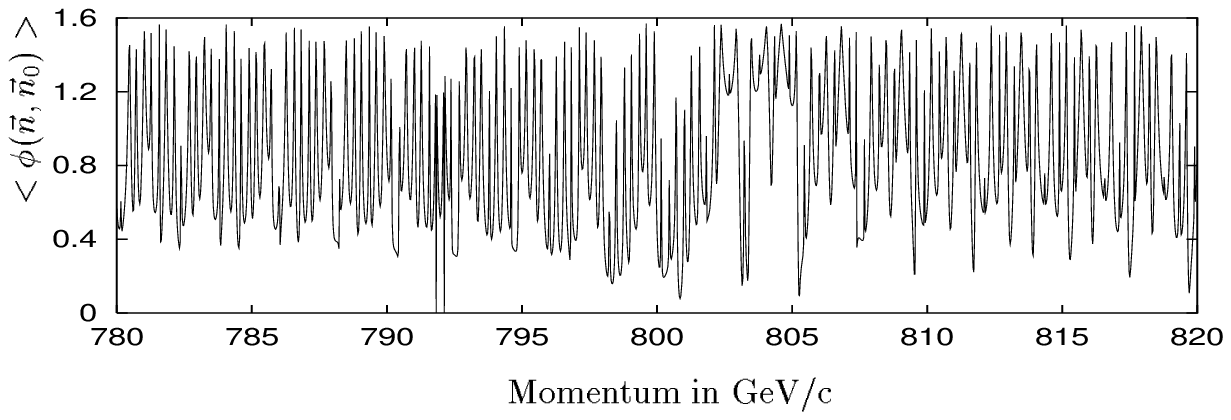}
\caption{Opening angle for DESY~III, PETRA, and HERA.
\label{fig:slim}}
\end{minipage}
\end{sidewaysfigure}

These opening angles are shown for DESY~III in figure
\ref{fig:slim}~(top) and it is apparent that at the places where
resonant spin perturbations are described by a large resonance
strength, the invariant spin field has a large opening angle.  It is
obvious when comparing with the resonance strength of figure
\ref{fig:res}~(top) that the influence of different resonances does
not overlap in the linearized spin approximation.  At PETRA energies
of up to 40GeV, the resonances already come very close to each other
as seen when comparing figure \ref{fig:slim}~(middle) with figure
\ref{fig:res}~(middle) and one can only barely expect an isolated resonance
approximation to lead to accurate results.  For high energies between
780 and 820GeV/c in HERA, figure \ref{fig:slim}~(bottom) clearly shows
that one cannot speak of isolated resonances.  Often the influences of
4 resonances overlap.

The approximation of linearized spin motion contains all first order
orbital frequencies, since it linearized the precession vector
$\vec\Omega$ with respect to $\vec z$.  However, in contrast to the
isolated resonance model, none of these resonances is ignored and the
effect of overlapping resonances can be seen.

It is possible to recover the first order isolated resonance strength
from the one turn spin orbit transport matrix.  In analogy to the
complex notation for the spin component perpendicular to $\vec n_0$,
the perturbing precession vector $\vec\omega$ is expressed by a
complex function $\omega(\vec z,\vec\theta)$ as
$\vec\omega=\Re\{\omega\}\vec m_0+\Im\{\omega\}\vec
l_0+(\vec\omega\cdot\vec n_0)\vec n_0$.  Inserting this into the
Thomas-BMT equation, one obtains
\begin{equation}
\alpha'=-i\sqrt{1-\vec\alpha^2}\omega+i\alpha(\vec\omega\cdot\vec
n_0)\ .
\end{equation}
In the case of spins which are nearly parallel to $\vec n_0$, one can
linearize in $\alpha$ and $\vec z$.  For a spin which was initially
parallel to $\vec n_0$ one obtains
$\alpha(\theta)\approx-i\int_0^\theta\omega d\theta\;$.  Comparing
with equation (\ref{eq:res}) one can express the resonance strength as
$\epsilon_{\nu_0}=i\lim_{N\to\infty}\frac{1}{2 \pi N}\alpha(2\pi N)\;$.  The
resonance strength can therefore be computed from $\underline
M_{77}^N/N$ for large $N$.  The computation becomes very efficient if
one uses $\underline M_{77}^{2N}=(M_{77}^N)^2$ iteratively.

The coordinate vectors $\vec m_0(2\pi)$ and $\vec l_0(2\pi)$ to which
$\alpha(2\pi)$ refers have rotated by the spin tune $\nu_0$, whereas
the final spin coordinate $\alpha_f$ computed by $\underline M_{77}$
refers to the coordinate vectors $\vec m_0(0)$ and $\vec l_0(0)$.
Therefore $\alpha(2\pi N)=\alpha_f\exp(-i2\pi N\nu_0)$, and
$\epsilon_{\nu_0}$ can be computed from powers of the one turn matrix,
which can most efficiently be evaluated in diagonal form,
\begin{eqnarray}
\epsilon_{\nu_0}
&=&
i\lim_{N\to\infty}\frac{1}{2\pi N}\alpha(2\pi N)
=
i\lim_{N\to\infty}\frac{1}{2\pi N}
(0,e^{-iN2\pi\nu_0})
{\underline M\ \ \ \ \underline 0\choose {\vec G}^T\ e^{i2\pi\nu_0}}^N
{\vec z\choose 0}\\
&=&
i\lim_{N\to\infty}
e^{-iN2\pi\nu_0}\frac{1}{2\pi N}\sum_{j=0}^{N-1}
[
e^{i(N-j-1)2\pi\nu_0}{\vec G}^T\underline A^{-1}\underline\Lambda^j]
\underline A\vec z\\
&=&
ie^{-i2\pi\nu_0}G_lA^{-1}_{lk}A_{km}z_m
\lim_{N\to\infty}\frac{1}{2\pi N}\sum_{j=0}^{N-1}e^{i 2\pi j(\nu_k-\nu_0)}
\end{eqnarray}
where one has to sum over equal indices $k$, $l$, and $m$.  This
formula shows that the resonance strength is always zero, except at a
resonance condition $\nu_0=m+\nu_k$. At such a spin tune, the
resonance strength is given by
\begin{equation}
2\pi|\epsilon_{\nu_0=\nu_k}|
=
|{\vec G}^T\underline A^{-1}\underline{\rm diag}(0...1...0)\underline
A\vec z|
=
|{\vec G}^T\underline A^{-1}(0...\sqrt{J_k}e^{i\Phi_k}...0)^T|
=
|\vec G\cdot\vec v_k|\sqrt{J_k}\ .
\end{equation}
The 1 in the diagonal matrix is in position $k$.  Here $\underline
A^{-1}(0...\sqrt{J_k}e^{i\Phi_k}...0)^T$ is the initial value
for a phase space trajectory which has only Fourier components with
frequencies $\nu_k$ plus integers and the $k$th eigenvector $\vec v_k$
of $\underline M$ has been used.  The infinite Fourier integral in
equation (\ref{eq:res}) has been reduced to the scalar product between
the bottom vector of $\underline M_{77}$ and an eigenvector of
$\underline M$.  This very simple formula is used in the program
SPRINT.

\subsection{Non-perturbative methods}

While one does not drop Fourier coefficients in the approximation of
linearized spin motion, there are other limitations.  The
approximation is no longer justified when $|n_\alpha|$ becomes large,
which happens close to resonances in the figures \ref{fig:slim}.
Therefore the validity of linearized spin motion had to be be checked
by computing the invariant spin field non-perturbatively.  In the last
few years two iterative higher order and three non-perturbative
methods of computing the invariant spin field have been developed
\cite{hoff98b}.  All of these methods agree within their ranges of
mutual applicability.  The invariant spin field obtained from a
non-perturbative method contains the effect of all Fourier
coefficients in $\vec\Omega$.  When comparing this spin field with
$\vec n_1$, it was found that linearized orbit motion describes the
opening angle and thus the maximum storeable polarization well in
domains where the opening angle is small.  At the critical energies,
where the maximum polarization is low during the acceleration process,
non-perturbative methods become essential for simulation and results
obtained with the computationally quick linearization of spin motion
should always be checked with more time consuming non-perturbative
methods if possible.

One application of this strategy is the filtering method
\cite{hoff96e99}.  Four or eight Siberian Snakes
are inserted into HERA to fix the spin tune to $1/2$ for all energies
and to let $\vec n_0$ be vertical in the flat arcs.  These conditions
do not fix all snake angles.  Currently there is, however, no
established formula to determine good snake angles.  Since the opening
angle of $\vec n(\vec z)$ is such a critical quantity for high energy
polarized proton acceleration, we have decided to maximize $<\vec n>$
by choosing snake angles.  A computer code was written which tested
approximately $10^6$ snake schemes and computational speed was
therefore essential.  Linearized spin motion was used to find the
8-snake-schemes with smallest average value of $|n_\alpha|$ over the
acceleration cycle.  These filtered snake schemes then also had
relatively small opening angles when computed non-perturbatively with
stroboscopic averaging \cite{hoff96d}.

Two other indications showed that this filtering leads to good snake
schemes.  (1) Tracking simulations of the complete ramp process showed
that the snake schemes found by filtering leads to less depolarization
\cite{hoff98d} than other schemes which were initially proposed. (2)
Computation of the amplitude dependent spin tune, which can only be
performed when $\vec n(\vec z)$ has been found non-perturbatively,
shows that snakes schemes found by filtering have significantly less
spin tune spread over orbital amplitudes than other proposed schemes
\cite{hoff99f}.  With the optimal scheme for four Siberian Snakes in
HERA it turned out to be possible to accelerate in computer
simulations approximately 65\% of the beam to high energy with little
loss of polarization as long as no closed orbit distortions were
present \cite{hoff99b1}.
In simulations, the current 1mm rms closed orbit distortions lead to
depolarization \cite{golubeva99}.  Therefore either the closed orbit
will have to be controlled more accurately or techniques which make
the spin motion less sensitive to closed orbit distortion
\cite{derbenev99} will have to be utilized.

\paragraph{Acknowledgment:}  I thank D.P.Barber for carefully reading
the manuscript.


\end{document}